\documentclass[11pt,twoside]{article}
\usepackage{asp2010}

\resetcounters

\newcommand{\aspconf}{ASP Conf.\ Ser.}

\begin{document}

\title{Significant problems in FITS limit its use in modern astronomical research}
\author{Brian~Thomas$^1$, Tim~Jenness$^2$, Frossie~Economou$^1$, Perry~Greenfield$^3$,
Paul~Hirst$^4$, 
David~S.~Berry$^5$, Erik~Bray$^2$,
Norman~Gray$^6$, Demitri~Muna$^7$, James~Turner$^8$,
Miguel~de~Val-Borro$^9$, Juande~Santander-Vela$^{10}$,
David~Shupe$^{11}$, John~Good$^{11}$, and G.~Bruce~Berriman$^{11}$
\affil{$^1$National Optical Astronomy Observatory, 950 N.\ Cherry Ave,
Tucson, AZ 85719, USA}
\affil{$^2$Department of Astronomy, Cornell University, Ithaca, NY 14853, USA}
\affil{$^3$Space Telescope Science Institute, 3700 San Martin Drive,
  Baltimore, MD 21218, USA}
\affil{$^4$Gemini Observatory, 670 N.\ A`oh\=ok\=u Place, Hilo, HI
  96720, USA}
\affil{$^5$Joint Astronomy Centre, 660 N.\ A`oh\=ok\=u Place, Hilo, HI
96720, USA}
\affil{$^6$School of Physics \& Astronomy, University of Glasgow,
  Glasgow, G12~8QQ, United Kingdom}
\affil{$^7$Department of Astronomy, Ohio State University, Columbus,
  OH 43210, USA}
\affil{$^8$Gemini Observatory, Casilla 603, La Serena, Chile}
\affil{$^9$Department of Astrophysical Sciences, Princeton University,
Princeton, NJ 08544, USA}
\affil{$^{10}$Instituto de Astrof\'{i}sica de Andaluc\'{i}a, Glorieta
  de la Astronom\'{i}a s/n, E-18008, Granada, Spain}
\affil{$^{11}$Infrared Processing and Analysis Center, Caltech,
  Pasadena, CA 91125, USA}
}

\begin{abstract}
  The Flexible Image Transport System (FITS) standard has been a great
  boon to astronomy, allowing observatories, scientists and the public
  to exchange astronomical information easily. The FITS standard is,
  however, showing its age. Developed in the late 1970s the FITS
  authors made a number of implementation choices for the format that,
  while common at the time, are now seen to limit its utility with modern
  data. The authors of the FITS standard could not appreciate the
  challenges which we would be facing today in astronomical
  computing. Difficulties we now face include, but are not limited to,
  having to address the need to handle an expanded range of
  specialized data product types (data models), being more conducive
  to the networked exchange and storage of data, handling very large
  datasets and the need to capture significantly more complex metadata
  and data relationships.

  There are members of the community today who find some (or all) of
  these limitations unworkable, and have decided to move ahead with
  storing data in other formats. This reaction should be taken as a
  wakeup call to the FITS community to make changes in the FITS
  standard, or to see its usage fall. In this paper we detail some
  selected important problems which exist within the FITS standard
  today.  It is not our intention to prescribe specific remedies to
  these issues; rather, we hope to call attention of the FITS and
  greater astronomical computing communities to these issues in the
  hopes that it will spur action to address them.
\end{abstract}

\section{Introduction}

The Flexible Image Transport System standard
\citep[FITS;][]{1981A&AS...44..363W,2001A&A...376..359H,2010A&A...524A..42P}
has been a fundamental part of astronomical computing for a
significant part of the past 4 decades. FITS has provided the central
means to store and exchange astronomical data and, because of hard
work of the FITS community, it has become a relatively easy exercise
for application writers, archivists and end user scientists to
interchange data and work productively on many computational astronomy
problems. 



While there have been significant changes, the FITS standard has
evolved very slowly since its creation in the late 1970s. FITS has
added new types of metadata conventions such as World Coordinate
System
\citep[WCS;][]{2002A&A...395.1061G,2002A&A...395.1077C,2006A&A...446..747G}
representation and data serializations such as variable length binary
tables \citep{1995A&AS..113..159C}. Nevertheless, these changes have
not been sufficient to match the greater evolution in astronomical
research over the same period of time.

Astronomical research now goes beyond the paradigm of the original
scientific team consuming only the observational data for which they
proposed. Astronomy researchers have shifted towards utilizing the
observations of others, accessing data from remote archives over the
Internet, and combining these data with the original observations (or
theoretical calculations) in order to obtain better and wider ranging
scientific results. Many research projects now involve many diverse
data sets from a range of sources and instruments in astronomy now
produce many orders of magnitude larger datasets than were common at
the time FITS was born.  
Additionally, astronomers have also come to increasingly rely on 
others to write software to help process and analyze their data. Common 
libraries, analysis environments, pipeline processed data, applications 
and services provided by third parties form a crucial foundation for many
astronomers' toolboxes. 

This evolution in research practice poses many new challenges for 
the 21\textsuperscript{st} century. The large volume of data, the shared 
software infrastructure, the distributed nature of the data holdings 
and the increasing complexity of the information we capture mandates that 
the data format used will enable the machine to do as much as possible
handle the interchange, storage and processing of scientific information. 

Because FITS has shown difficulties in these areas some members of the community 
have gravitated away from FITS seeking more capable solutions.
Other data formats serving in this role include the Starlink Hierarchical Data System
\citep[HDS;][]{1982QJRAS..23..485D,P91_adassxxiii} and the adoption of the Hierarchical 
Data Format version 5 (HDF5) by the Low-Frequency Array for radio astronomy 
project \citep[LOFAR;][]{2011ASPC..442...53A}.
We predict that the use of FITS will inevitably decline should it not adapt 
to these new challenges.

In this paper we detail some selected important problems which exist 
within the FITS standard today.  It is not our intention to prescribe 
specific remedies to these issues; rather, we hope to call attention 
of the FITS and greater astronomical computing communities to these 
issues in the hopes that it will spur action to address them.

\section{Problems -- Deficiencies of FITS for Modern Astronomical Computing}

There is not enough space in this paper to go into a detailed
description of the deficiencies that we see are present in the current
incarnation of FITS. Instead we will summarize the issues and present
a more detailed examination in a subsequent paper.

\subsection{Lack of versioning, semantics and encodings}

There is no standard way to specify the version of a given FITS file or
what extensions it supports. You must read the file and determine
dynamically which extensions are present and whether they are
understood. The ``once FITS, forever FITS'' maxim gives you some
confidence that you can always read a FITS file but you cannot be sure
if there is something that you do not understand.  Explicit versioning
of FITS files will help but there also needs to be a way to declare that
a particular data model is being used from variants such as OIFITS
\citep{2006SPIE.6268E.106T}, MBFITS \citep{2006A&A...454L..25M}, and
FITS-IDI \citep{2011AIPS114}, and to validate the contents against a
namespaced schema.

The allowed character set in FITS of 7-bit US-ASCII is overly restrictive in a
Unicode world. It is unacceptable that FITS authors cannot use special
scientific or mathematical symbols (e.g.\ a degree symbol) or capture
non-English text in tables or FITS headers.

\subsection{Missing data models}

FITS can support basic data models such as tables and multi-dimensional 
images but lacks many higher level data models which enable scientific
data description. To start with, there is no standardized way of associating
the basic models in a related manner. Determining that a particular image 
extension contains the variance or mask for another image relies on string 
parsing and shared convention. 

Ironically, for a format designed to handle astronomy data, FITS lacks shared
models which describe scientific errors or data quality. Archiving is
a primary use case for FITS however it lacks a sufficiently rich model 
for capturing the history/provenance of the data.
The HISTORY keyword provides just a textual representation of
provenance which cannot be machine-read, and with a very loose 
meaning outside particular applications.

Finally, the current FITS WCS data models are complex yet incomplete and
inflexible. There needs to be a way to stack mappings in an arbitrary
manner to allow for flexible model development \citep[see
e.g.][]{1998ASPC..145...41W,2012ASPC..461..825B,O35_adassxxii}.

\subsection{Inflexibility in representing metadata and data}

The 80-character card image drives a number of subsequent limitations which
result in poor metadata description (8-character keyword, 68-character limit in
keyword values, and cumbersome CONTINUE card constructs).  This out-dated
restriction also results in the awkward implementation of some conventions, such
as ESO HIERARCH \citep{2009Wic}, that can not overcome the underlying
limitations of representation. Additionally, the lack of namespaces results in
uncertainty over metadata meaning with other FITS files. Finally, the 2880
record is a minor but annoying restriction which results in wasteful blocks of
whitespace in many FITS files, hampers the use of FITS to capture very small,
but richly described data, and impedes the real-time writing of FITS files.

\subsection{Inadequate support for large, distributed data}

Modern data sets can result in files of several terabytes that must be
distributed across multiple file systems. The FITS grouping convention
tries to provide this facility but is neither robust nor transparent
enough. Additionally, streaming indeterminately sized data sets
to files must be supported.

\section{Summary -- Significant Problems exist in the FITS standard}

The problems which we have described are real and significant. We
do not wish to recommend a particular solution here. Action to correct
these issues should flow from constructive community discussion, and
offered solutions to these problems. Possible solutions may involve
moving existing FITS conventions into the core standard,
modification of the FITS standard to remove limitations or possibly
transferring the FITS data model over into a new serialization, or
some selection of these actions.

These technical problems will be solved one way or another. If the
community is not willing to do the hard work of hammering out a
universal (or widely-adopted) approach, individual projects will
continue to make their own ad-hoc solutions. Data formats will become
increasingly fragmented and we will no longer enjoy the easy
interoperability that FITS has provided for many years.

\bibliography{P90}

\end{document}